\documentclass[reprint,aps,prl]{revtex4-2}

\usepackage[pdftex]{graphicx}
\usepackage{amsmath}

\begin{document}

\title{Persistence of charge density wave fluctuations in the absence of long-range order in a hole-doped kagome metal}

\author{Terawit Kongruengkit}
\affiliation{Materials Department, University of California, Santa Barbara, California 93106, USA}

\author{Andrea N. Capa Salinas}
\affiliation{Materials Department, University of California, Santa Barbara, California 93106, USA}

\author{Ganesh Pokharel}
\affiliation{Materials Department, University of California, Santa Barbara, California 93106, USA}

\author{Brenden R. Ortiz}
\affiliation{Materials Department, University of California, Santa Barbara, California 93106, USA}

\author{Stephen D. Wilson}
\affiliation{Materials Department, University of California, Santa Barbara, California 93106, USA}

\author{John W. Harter}
\email[Corresponding author: ]{harter@ucsb.edu}
\affiliation{Materials Department, University of California, Santa Barbara, California 93106, USA}

\date{\today}

\begin{abstract}
The kagome metals $A$V$_3$Sb$_5$ ($A$~=~K, Rb, Cs) exhibit a complex interplay between charge density wave (CDW) order and superconductivity. In this study, we use ultrafast coherent phonon spectroscopy to probe the evolution of CDW order in hole-doped CsV$_3$Sb$_{5-x}$Sn$_x$ across a broad range of compositions ($0 \leq x \leq 0.68$). While thermodynamic and diffraction measurements show long-range CDW order vanishes above $x \approx 0.05$, we observe persistent signatures of CDW fluctuations up to the highest doping levels, with correlation times on the order of several picoseconds. These results indicate the presence of robust fluctuating charge order that survives well beyond the established CDW phase boundary. Furthermore, these fluctuations are enhanced near a doping-tuned quantum phase transition at $x^* \approx 0.15$, which coincides with a local minimum in the superconducting $T_\mathrm{c}$ double-dome. Additional measurements on Ti- and K-substituted samples confirm that this behavior is intrinsic to hole doping and not tied to disorder. Overall, our findings suggest that CDW fluctuations play a central role in the electronic phase diagram of $A$V$_3$Sb$_5$ and may mediate or compete with superconductivity.
\end{abstract}

\maketitle

The recent discovery of the $A$V$_3$Sb$_5$ ($A$~=~K, Rb, Cs) material family established a new arena for testing theoretical predictions of the diverse electronic instabilities of the kagome lattice~\cite{guo2009,ko2009,yu2012,kiesel2013,wang2013,mazin2014}. These compounds, crystallizing in the $P6/mmm$ space group, feature perfect quasi-two-dimensional kagome lattices of vanadium atoms that host a range of exotic electronic phenomena, most notably a complex charge density wave (CDW) phase below $T_\mathrm{CDW} \sim 100$~K out of which superconductivity emerges below $T_\mathrm{c} \sim 1$~K~\cite{ortiz2019,ortiz2020,ortiz2021a,yin2021}. Upon entering the CDW phase, the crystal structure undergoes a three-dimensional distortion, expanding the unit cell. Experimental and theoretical studies indicate that the CDW in KV$_3$Sb$_5$ and RbV$_3$Sb$_5$ is captured by a $2\times2\times2$ supercell, while CsV$_3$Sb$_5$ exhibits unusual competition between $2\times2\times2$ and $2\times2\times4$ superstructures~\cite{ratcliff2021,liang2021,miao2021,ortiz2021b,kautzsch2023a}. Among the numerous studies of the CsV$_3$Sb$_5$ system, many have focused on doping and its influence on the CDW and superconducting phases~\cite{oey2022,ding2022,liu2022a,yang2022,kautzsch2023b,capasalinas2023}. Light hole doping via tin substitution for antimony reveals a direct competition between the CDW state and superconductivity, yielding a phase diagram with a rapidly suppressed $T_\mathrm{CDW}$ and a correspondingly enhanced $T_\mathrm{c}$. Notably, for higher doping levels, the superconducting phase shows double-dome behavior, with the local minimum $T_\mathrm{c}$ appearing near the doping where $T_\mathrm{CDW}$ extrapolates to zero~\cite{oey2022}.

Recent x-ray diffraction experiments suggest that an incommensurate charge ordered state emerges following the complete suppression of long-range CDW order by hole doping~\cite{kautzsch2023b}. The detailed progression from long-range CDW order to short-range charge correlations in $A$V$_3$Sb$_5$ and its influence, if any, on the superconducting phase is still an open question, however. In this Letter, we investigate the evolution of charge order in CsV$_3$Sb$_{5-x}$Sn$_x$ over a wide range of doping compositions from $x = 0$ to $x = 0.68$ using ultrafast coherent phonon spectroscopy. Despite the CDW transition being essentially undetectable above $x \approx 0.05$ by thermodynamic and diffraction probes~\cite{oey2022,kautzsch2023b}, our measurements reveal signatures of fluctuating CDW order up to at least $x = 0.68$. We thus uncover the conspicuous persistence of CDW correlations even in the absence of long-range order, at temperatures and doping levels far outside the accepted CDW phase boundary. We also investigate CsV$_{3-y}$Ti$_y$Sb$_5$ and Cs$_{1-z}$K$_z$V$_3$Sb$_5$ and demonstrate that the specific mechanism of doping and the presence or absence of disorder are irrelevant to the behavior that we uncover. At low temperatures, we identify a doping-tuned quantum phase transition (QPT) near which fluctuations are especially large. The location of this QPT appears to coincide with the local minimum of the superconducting $T_\mathrm{c}$ double-dome. Taken as a whole, our results support a picture of persistent, ubiquitous, and strong CDW fluctuations upon hole doping of the $A$V$_3$Sb$_5$ material family even at low temperatures, markedly influencing the superconducting phase.

\begin{figure*}[t]
\includegraphics{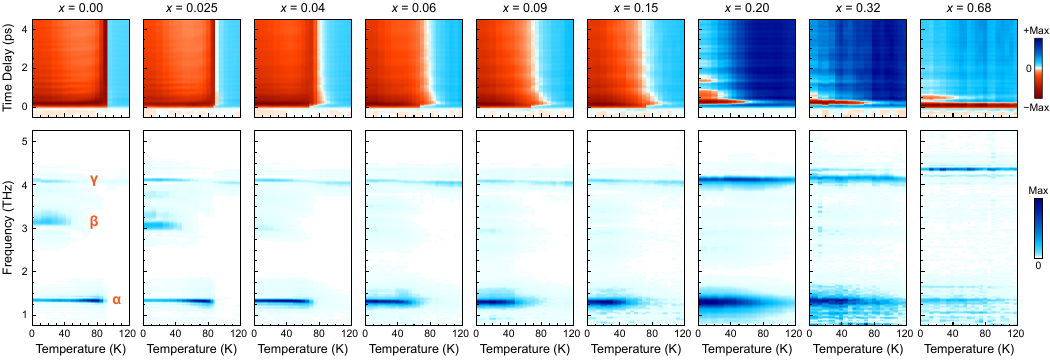}
\caption{\label{figure1} Transient reflectivity and coherent phonon spectroscopy for CsV$_3$Sb$_{5-x}$Sn$_x$. Sample composition ranges from $x = 0$ on the left to $x = 0.68$ on the right. The top row shows the transient change in reflectivity ($\Delta R/R$). For lightly doped samples with a well-defined transition, above $T_\mathrm{CDW}$ the reflectivity increases after the pump pulse (blue regions), while below $T_\mathrm{CDW}$ it decreases (red regions). At intermediate doping levels, where long-range order is suppressed, the sharp transition is replaced by a more gradual change from positive to negative. At high doping levels, $\Delta R/R$ is positive for all temperatures. The bottom row shows temperature--frequency maps of the magnitude of the Fourier transform of coherent phonon oscillations induced by the pump pulse, obtained after subtraction of an exponential background from $\Delta R/R$. Following Ref.~\citenum{ratcliff2021}, we label the spectral features $\alpha$~(1.3~THz), $\beta$~($\sim$3~THz), and $\gamma$~(4.1~THz). The $\gamma$ mode is a fully-symmetric phonon active at all temperatures. In contrast, the $\alpha$ mode only becomes active through CDW order or its fluctuations.}
\end{figure*}

Single crystals of CsV$_3$Sb$_{5-x}$Sn$_x$, CsV$_{3-y}$Ti$_y$Sb$_5$, and Cs$_{1-z}$K$_z$V$_3$Sb$_5$ were synthesized by a self-flux method, as detailed in Refs.~\citenum{kautzsch2023b}, \citenum{pokharel2025}, and \citenum{ortiz2023}, respectively. The compositions of crystal batches were confirmed by energy-dispersive x-ray spectroscopy to 1\% accuracy. Time-resolved optical reflectivity measurements were performed on freshly-cleaved (001) surfaces of crystals mounted in an optical cryostat. A non-collinear optical parametric amplifier was used to generate $\sim$70~fs signal~(800~nm) and idler~(1515~nm) pulses at a repetition rate of 50~kHz, which were used as probe and pump beams, respectively. Both pulses had fluences of $\sim$100~$\mu$J/cm$^2$ and were linearly polarized in-plane. Scans of reflectivity as a function of pump-probe time delay were acquired by synchronizing the signal from a silicon photodiode with the position of a rapidly oscillating optical delay stage, with each resulting scan representing an average of over $10^5$ sweeps of the stage.

Coherent phonon spectroscopy has been shown to be a particularly sensitive probe of the CDW order in $A$V$_3$Sb$_5$~\cite{ratcliff2021,wang2021,deng2025a,deng2025b}. In this technique, ultrafast optical pump pulses coherently excite phonon oscillations via the displacive excitation of coherent phonon (DECP) mechanism~\cite{zeiger1992}. In the DECP mechanism, absorption of the pump photons instantaneously excites electrons, abruptly altering the system's free energy. This shifts the quasi-equilibrium positions of the atoms within the unit cell, resulting in a restoring force that initiates the coherent phonon motion. The amplitudes of the decaying oscillations are subsequently measured through their coupling to the reflectivity of a second probe pulse. A key feature of this process is that only fully-symmetric phonons at the $\Gamma$ point are excited by the pump, which can be used to infer details about symmetry breaking by the CDW. The time-dependent relative change in reflectivity, denoted $\Delta R/R$, upon which the phonon oscillations are superposed, can itself also reveal information about the CDW because band renormalization, partial gapping of the Fermi surface, and spectral weight transfer will influence the pump-induced change in reflectivity~\cite{liu2021,wenzel2022}.

Figure~\ref{figure1} shows raw transient reflectivity data and Fourier transforms of corresponding phonon oscillations for a series of CsV$_3$Sb$_{5-x}$Sn$_x$ samples with doping values ranging from $x = 0$ to $x = 0.68$. In the transient reflectivity, we observe a striking change from positive to negative $\Delta R/R$ upon cooling through $T_\mathrm{CDW}$. This change in the sign of $\Delta R/R$ is quite abrupt for light doping levels, becomes broadened for intermediate doping levels, and completely vanishes above $x \approx 0.15$. The reflectivity is a complex function of the transiently-populated electronic states in the material and their subsequent relaxation, and as discussed above, the change in the sign of $\Delta R/R$ is the result of marked changes in these electronic states induced by the CDW~\cite{liu2021,wenzel2022}. In the coherent phonon spectra, obtained from $\Delta R/R$ by subtracting the slow exponential recovery and Fourier transforming the surviving oscillatory component, we observe three main spectral features. Following Ref.~\citenum{ratcliff2021}, we label these $\alpha$, $\beta$, and $\gamma$. The $\gamma$ mode at 4.1~THz is a fully-symmetric phonon that is active at all temperatures and is related to a breathing motion of the out-of-plane antimony atoms towards and away from the vanadium kagome planes~\cite{ratcliff2021}. The $\beta$ spectral feature at $\sim$3~THz likely comprises a pair of phonon modes that become active due to $C_6 \rightarrow C_2$ rotational symmetry breaking by the interlayer $\pi$ phase shift of the CDW~\cite{deng2025a,deng2025b}. This feature, which was originally thought to be associated with the CDW amplitude mode, appears only far below $T_\mathrm{CDW}$ because it is significantly overdamped at higher temperatures, as observed by Raman spectroscopy~\cite{liu2022b,he2024}. Finally, the $\alpha$ mode at 1.3~THz represents a phonon at the $L$ point associated with out-of-plane motion of the cesium atoms~\cite{ratcliff2021}. Below $T_\mathrm{CDW}$, this phonon becomes fully-symmetric because of its coupling to the CDW order parameter, which folds the $L$ point to $\Gamma$ by enlarging the unit cell. Crucially, this makes the phonon appear in the coherent phonon spectrum abruptly below $T_\mathrm{CDW}$. The $\alpha$ mode, together with the sign of $\Delta R/R$, represent key experimental features that we will focus on for the remainder of this Letter.

\begin{figure}[t]
\includegraphics{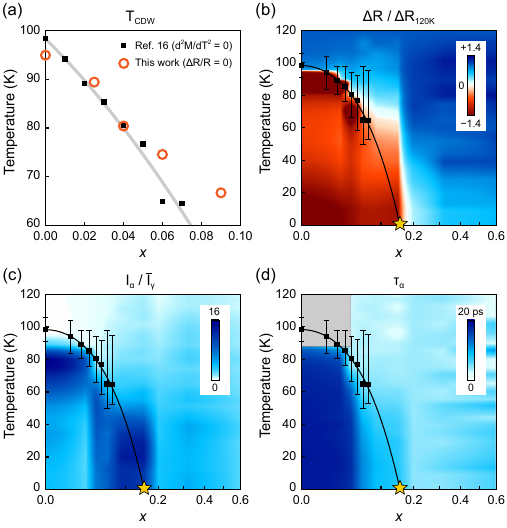}
\caption{\label{figure2} Doping--temperature phase diagram. (a)~$T_\mathrm{CDW}$ extracted from magnetization data in Ref.~\citenum{oey2022} (black squares) and from transient reflectivity data in this work (orange circles). The gray line is a least-squares quadratic polynomial fit to the magnetization data. (b)~Map of the transient reflectivity change $\Delta R$ normalized by the change at 120~K. (c)~Map of the $\alpha$ mode intensity normalized by the average $\gamma$ mode intensity. (d)~Lifetime of the $\alpha$ mode. The gray region shows where the mode is too weak to extract a meaningful lifetime. In panels (b)-(d), maps are generated via bilinear interpolation, black squares show $T_\mathrm{CDW}$ from magnetization, black lines are quadratic fits, error bars show the full width at half maximum of the $dM/dT$ peak, and stars at $T = 0$ illustrate the QPT at $x^*$.}
\end{figure}

We define three distinct experimental quantities to characterize the CDW order across doping--temperature phase space: the average change in reflectivity between 10 and 15~ps, normalized by that at 120~K ($\Delta R/\Delta R_\mathrm{120K}$), which captures CDW-induced changes in the electronic structure, specifically through a change in sign; the integrated spectral weight of the $\alpha$ mode, normalized by the temperature-averaged weight of the $\gamma$ mode ($I_\alpha/\bar{I}_\gamma$), which captures symmetry breaking by the CDW order parameter; and the lifetime of the $\alpha$ mode oscillation, defined as the reciprocal of the full width at half maximum of the spectral peak ($\tau_\alpha$), which captures decoherence of phonon oscillations by CDW fluctuations. Figure~\ref{figure2} shows doping--temperature maps of these three quantities.

The temperature at which $\Delta R/\Delta R_\mathrm{120K}$ changes sign can be used to define $T_\mathrm{CDW}$ for each doping level. In Fig.~\ref{figure2}(a), we plot these values on top of those extracted from magnetization data in Ref.~\citenum{oey2022}, where $T_\mathrm{CDW}$ was defined as the location of the peak in $dM/dT$ curves. We find almost perfect agreement between the two independent methods. Notably, however, they start to deviate upon approaching moderate doping levels, where the $dM/dT$ curves become significantly broadened and no longer exhibit sharp peaks, making it challenging to infer $T_\mathrm{CDW}$~\cite{oey2022}. Nevertheless, by fitting the magnetization data to a quadratic polynomial, we can extrapolate $T_\mathrm{CDW}$ to higher doping levels. Such an extrapolation leads to the prediction of a QPT, where $T_\mathrm{CDW} \rightarrow 0$ at a critical doping $x^* \approx 0.15$. As Fig.~\ref{figure2}(b) shows, this prediction is accurately reflected in the $\Delta R/\Delta R_\mathrm{120K}$ map, which at low temperatures is negative below $x = 0.15$ and positive above it. Outside the extrapolated phase boundary and above the QPT, however, we find a region of phase space where $\Delta R/\Delta R_\mathrm{120K}$ remains negative. Instead, the phase boundary derived from our measurements appears to flatten out before abruptly terminating at $x^*$.

The spectral weight of the $\alpha$ mode, shown in Fig.~\ref{figure2}(c), offers a complementary view of the phase diagram: While $\Delta R/\Delta R_\mathrm{120K}$ is sensitive to CDW-induced changes in electronic states, $I_\alpha/\bar{I}_\gamma$ directly probes the translational symmetry breaking of the CDW via the folding of an $L$ point phonon to $\Gamma$~\cite{ratcliff2021}. Here we observe several remarkable features in the data. First, at light doping levels, we find the spectral weight is strongest right below $T_\mathrm{CDW}$, becoming weaker as temperature is lowered. Above $T_\mathrm{CDW}$, in contrast, the weight is essentially zero. Second, for doping levels approaching $x^*$, we find that this strong spectral weight has shifted to low temperatures, comprising a dome-shaped region near the QPT that includes the region of phase space where $\Delta R/\Delta R_\mathrm{120K}$ is negative. Third, for most of phase space outside the CDW phase boundary, we detect finite spectral weight. This is striking because one does not expect translational symmetry breaking outside the CDW phase boundary, and hence the $\alpha$ mode should remain silent (which is indeed the case for light doping levels).

To reconcile the observation of finite $I_\alpha$ with the absence of long-range order, at temperatures and doping levels far outside the CDW phase boundary, we must examine the coupling between the CDW order, the $\alpha$ mode, and the reflectivity. First, let us consider the CDW and the $\alpha$ mode amplitude $\xi_\alpha$. The ``3Q'' $2\times2\times2$ CDW in $A$V$_3$Sb$_5$ arises from the simultaneous condensation of three charge density waves with amplitudes $\phi_1$, $\phi_2$, and $\phi_3$ at the $M_1$, $L_2$, and $L_3$ points of the Brillouin zone, respectively, following a so-called ``MLL'' configuration that captures in-plane tri-hexagonal order with an interlayer $\pi$ phase shift~\cite{ratcliff2021}. The symmetries and momenta of the order parameters are such that a cubic coupling term of the form $\phi_1\phi_2\phi_3$ is allowed in the Landau free energy (this coupling term, for example, lifts the degeneracy of the tri-hexagonal and Star-of-David distortions~\cite{tan2021}). Because the $\alpha$ mode shares the same momentum and irreducible representation as one of the $L$ point orders, say $\phi_3$, a coupling of the form $\phi_3\xi_\alpha$ is also allowed. We obtain the free energy expansion $F(\xi_\alpha) = (a/2)\xi_\alpha^2 - g\phi_3\xi_\alpha + \cdots$, where $a$ and $g$ are constants, whose minimization yields $\xi_\alpha = (g/a)\phi_3$. This offers a straightforward microscopic mechanism for the DECP process: pump pulses transiently suppress charge order [$\phi_3 \rightarrow (1 - \delta)\phi_3$], for example through instantaneous heating or the excitation of electron--hole pairs, which effectively displaces the $\alpha$ mode by $\Delta\xi_\alpha = \delta(g/a)\phi_3$ from its new equilibrium position. Examining now the coupling of these order parameters to reflectivity, again by symmetry one finds $\Delta R = r\phi_3\Delta\xi_\alpha + \cdots$, where $\Delta R$ is the initial amplitude of reflectivity oscillations (directly proportional to $I_\alpha$) and $r$ is a constant. Combining the above equations, we arrive at a key result: $\Delta R = \delta(rg/a)\phi_3^2$. Thus, we have shown that $I_\alpha$ is proportional to the \textit{square} of $\phi_3$, and even in the absence of long-range order, where $\left\langle\phi_3\right\rangle = 0$, fluctuations of the CDW characterized by $\left\langle\phi_3^2\right\rangle > 0$ can lead to finite $I_\alpha$.

\begin{figure}[t]
\includegraphics{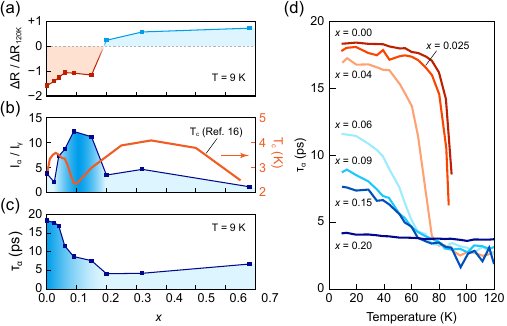}
\caption{\label{figure3} Doping dependence at low temperatures and evolution of the $\alpha$ mode lifetime. (a)~Low temperature transient reflectivity change $\Delta R/\Delta R_\mathrm{120K}$. (b)~$\alpha$ mode intensity normalized by the average $\gamma$ mode intensity (left axis) and superconducting onset $T_\mathrm{c}$ from Ref.~\citenum{oey2022} (right axis). The region of large $I_\alpha$ (signaling strong CDW fluctuations) coincides with the local minimum in the superconducting double-dome. (c)~Lifetime of the $\alpha$ mode. Approaching the QPT, where fluctuations grow large, $\tau_\alpha$ is correspondingly reduced. (d)~Lifetime of the $\alpha$ mode versus temperature for select doping levels, showing a reduced $\tau_\alpha$ near $T_\mathrm{CDW}$ even at light doping levels.}
\end{figure}

To test this conclusion, we map in Fig.~\ref{figure2}(d) the lifetime of the $\alpha$ mode oscillations. For light doping levels below $T_\mathrm{CDW}$, the oscillations have a relatively long temperature-independent lifetime of $\sim$20~ps. For moderate doping levels but still well within the CDW phase boundary, however, we find $\tau_\alpha$ is significantly reduced from this value. The region of shortened lifetime coincides almost perfectly with the dome of enhanced $I_\alpha$ near the QPT. Well outside the phase boundary at moderate to high doping levels, we find a relatively uniform short lifetime of $\sim$4~ps. The relationship between $I_\alpha$ and $\tau_\alpha$ at low temperatures is shown more clearly in Fig.~\ref{figure3}. Additionally, as Fig.~\ref{figure3}(d) demonstrates, we also find a reduced $\tau_\alpha$ very close to $T_\mathrm{CDW}$, even for light doping levels. Overall, our observations support a scenario whereby CDW fluctuations cause decoherence of the $\alpha$ mode oscillations, reducing $\tau_\alpha$. These fluctuations are largest (and $I_\alpha$ most intense) either just below $T_\mathrm{CDW}$, where critical thermal fluctuations are significant, or near the QPT, where quantum fluctuations may become relevant~\cite{vojta2003}. Indeed, recent ab initio calculations suggest that quantum zero-point motion could play an important role in the lattice dynamics of CsV$_3$Sb$_5$ despite the relatively heavy elements involved~\cite{chen2024}. We can estimate the correlation time of CDW fluctuations using the equation $1/\tau_\alpha = 1/\tau_0 + 1/\tau_\mathrm{CDW}$, where $\tau_0 \approx 20$~ps is the lifetime in the absence of fluctuations. Using $\tau_\alpha \approx 8$~ps, we find a CDW correlation time of $\tau_\mathrm{CDW} \approx 13$~ps in the region near the QPT. Far outside the CDW phase boundary, at high temperatures and doping levels, we find $\tau_\mathrm{CDW} \approx 5$~ps.

\begin{figure}[t]
\includegraphics{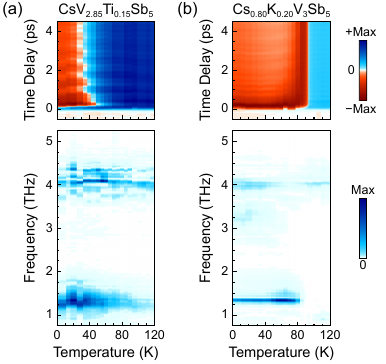}
\caption{\label{figure4} Transient reflectivity and coherent phonon spectroscopy for (a) CsV$_{2.85}$Ti$_{0.15}$Sb$_5$ and (b) Cs$_{0.80}$K$_{0.20}$V$_3$Sb$_5$. The top panel shows $\Delta R/R$ and the bottom panel shows the coherent phonon spectrum. Despite the different forms of disorder introduced by chemical substitution, measurements of the titanium-doped compound and potassium alloy are nearly identical to those of the tin-doped and undoped compounds, respectively.}
\end{figure}

Chemical doping unavoidably introduces crystallographic disorder, and it is not immediately clear to what extent such disorder may contribute to the ubiquitous CDW fluctuations that we have uncovered throughout the doping--temperature phase plane of CsV$_3$Sb$_{5-x}$Sn$_x$. To resolve this question, we have measured two other disordered compounds: CsV$_{3-y}$Ti$_y$Sb$_5$ ($y = 0.15$) and Cs$_{1-z}$K$_z$V$_3$Sb$_5$ ($z = 0.20$). The first material is hole doped via the substitution of titanium for vanadium, introducing a markedly different type of disorder into the crystal structure that may be expected to have a much stronger effect on the charge order (which arises directly from the vanadium valence electrons). The second material, in contrast, is undoped, with isovalent disorder introduced on the alkali metal site. Figure~\ref{figure4} shows transient reflectivity and coherent phonon spectra for the two compounds. Despite the different forms of disorder, the measurements are largely consistent with the tin series. In particular, the hole-doped titanium compound shows a broadened $\Delta R/R$ sign change and $\alpha$ mode spectral weight persisting to high temperatures, consistent with significant fluctuations of CDW order. In contrast, the potassium alloy is almost identical to undoped CsV$_3$Sb$_5$, demonstrating that introducing disorder while keeping the charge carrier concentration fixed does not increase CDW fluctuations. Incidentally, the coherent phonon spectrum of the titanium-doped compound appears significantly noisier than that of the other materials, especially with regard to the $\gamma$ mode, and we speculate that mesoscale chemical inhomogeneity could be more significant in this material. This is fully consistent with prior work, which found a stronger disorder potential in the titanium-doped compounds~\cite{pokharel2025}. The much larger ``mass disorder'' introduced by the doping ($m_\mathrm{V}/m_\mathrm{Ti} = 1.064$ while $m_\mathrm{Sb}/m_\mathrm{Sn} = 1.026$) could also play a role in the phonon dynamics.

In conclusion, our results establish that CDW fluctuations in CsV$_3$Sb$_5$ persist far beyond the disappearance of long-range order, spanning a broad region of the doping--temperature phase diagram. These fluctuations have correlation times on the order of several picoseconds. We also uncover evidence of a doping-tuned QPT near $x^* \approx 0.15$ that is marked by a dome of enhanced fluctuations. Notably, the abrupt termination of long-range CDW order at $x^*$ suggests a first-order rather than second-order QPT. This distinction has important ramifications. Whereas a second-order transition would allow quantum critical fluctuations to grow smoothly and diverge at the critical point, a first-order transition may involve phase coexistence, metastability, or abrupt symmetry changes, all of which can lead to complex dynamics and spatial inhomogeneities near the QPT~\cite{vojta2003}. Reports of a $2\times2\times4$ CDW superstructure in pristine CsV$_3$Sb$_5$, as opposed to the simpler $2\times2\times2$ order seen in KV$_3$Sb$_5$ and RbV$_3$Sb$_5$, may be intimately connected to the strong CDW fluctuations. A longer modulation period along the $c$ axis implies a more delicate balance of interlayer coupling and structural energetics, potentially rendering the system more susceptible to fluctuations. This sensitivity could help explain the pronounced fluctuation signatures we detect just below $T_\mathrm{CDW}$ at light doping levels, precisely where the $2\times2\times4$ superstructure is observed~\cite{kautzsch2023a}, as the system navigates between competing stacking orders.

We emphasize that the ubiquitous CDW fluctuations we have uncovered may not simply be a precursor to long-range CDW order, but could instead represent an intrinsically fluctuating phase that survives deep into the disordered regime. This picture is consistent with reports of incommensurate charge order at high doping levels~\cite{kautzsch2023b} and a so-called ``hidden order'' recently detected by quasiparticle interference~\cite{huang2025}. The gradual evolution with doping from commensurate to fluctuating and ultimately to incommensurate states suggests a continuous transformation of charge order, possibly governed by a frustrated energy landscape that becomes increasingly complex as the system is tuned by hole doping. These residual charge correlations may in turn influence the superconducting state~\cite{holbaek2025}. In particular, the coincidence of enhanced fluctuations near the QPT with the local minimum in the double-dome superconducting $T_\mathrm{c}$ profile [Fig.~\ref{figure3}(b)] clearly points to a nontrivial interplay. For example, CDW fluctuations could act to enhance superconductivity in certain regimes via fluctuation-mediated pairing, while suppressing it near the QPT due to competing order or decoherence. Altogether, our findings underscore the central role of CDW fluctuations in shaping the electronic phase diagram of the $A$V$_3$Sb$_5$ kagome family and invite further theoretical and experimental investigation into how these fluctuations couple to superconductivity and other emergent phenomena.

\section*{Acknowledgments}

This work was supported by the U.S. Air Force Office of Scientific Research (AFOSR) under Award No.~FA9550-22-1-0270. Sample synthesis was supported by the National Science Foundation (NSF) through Enabling Quantum Leap: Convergent Accelerated Discovery Foundries for Quantum Materials Science, Engineering, and Information (Q-AMASE-i): Quantum Foundry at UC Santa Barbara (DMR-1906325).

\end{document}